\documentclass[sn-mathphys]{sn-jnl}

\jyear{2021}%

\theoremstyle{thmstyleone}%
\theoremstyle{thmstyletwo}%

\theoremstyle{thmstylethree}%

\usepackage{lineno,hyperref}

\usepackage{graphicx}
\usepackage{caption}
\usepackage{dcolumn}
\usepackage{bm}
\usepackage{mathtools}
\usepackage{amsmath, amsthm}
\usepackage{xcolor}
\usepackage{multirow}
\usepackage{float}
\begin{document}

\title[Machine learning interatomic potential for high-throughput screening and optimization of high-entropy alloys]{Machine learning interatomic potential for high-throughput screening and optimization of high-entropy alloys}


\author*[1]{\fnm{Anup} \sur{Pandey}}\email{anup@lanl.gov}

\author[2]{\fnm{Jonathan} \sur{Gigax}}\email{jgigax@lanl.gov}

\author[1]{\fnm{Reeju} \sur{Pokharel}}\email{reeju@lanl.gov}
\affil[1]{\orgdiv{Material Science and Technology}, \orgname{Los Alamos National Laboratory}, \orgaddress{\city{Los Alamos}, \state{NM} \postcode{87545}, \country{USA}}}

\affil[2]{\orgdiv{Center for Integrated Nanotechnologies}, \orgname{Los Alamos National Laboratory}, \orgaddress{\city{Los Alamos}, \state{NM} \postcode{87545}, \country{USA}}}

\abstract {We have developed a machine learning-based interatomic potential (MLIP) for the quaternary MoNbTaW (R4) and quinary MoNbTaTiW (R5) high-entropy alloys (HEAs). MLIPs enabled accurate high-throughput calculations of elastic and mechanical properties of various non-equimolar R4 and R5 alloys, which are otherwise very time-consuming calculations when performed using density functional theory (DFT).  We demonstrate that the MLIP  predicted properties compare well with the  DFT results on various test cases and are consistent with the available experimental data. The MLIPs are also utilized for high-throughput optimization of non-equimolar R4 candidates by guided iterative tuning of R4 compositions to discover candidate materials with promising hardness-ductility combinations. We also used this approach to study the effect of Ti concentration on the elastic and mechanical properties of R4, by statistically averaging the properties of over 100 random structures. MLIP predicted hardness and bulk modulus of equimolar R4 and R5 HEAs are validated using experimentally measured Vicker's hardness and modulus. This approach opens a new avenue for employing MLIPs for HEA candidate optimization. }




\maketitle

\section{Introduction}\label{sec1}

Alloy development is a costly, multi-year exercise.  Selection of elements that produce a material with desired properties requires careful consideration of multi-component interactions.  Further compounding on this difficulty is the perpetual need to optimize existing alloys for different applications, such as those for high temperature or radiation environments. The development of a computationally-based methods would serve to significantly reduce the exploratory cost when designing a new alloy or optimizing existing alloys.  

Highly accurate first principles calculations are limited by the simulated length and time scale. Therefore, it is practically impossible to explore the vast configurational space of multi-component alloys  (MCAs) using computationally demanding \emph{ab initio} methods such as density functional theory (DFT). On the other hand, molecular dynamics (MD) simulations based on empirical force fields are much faster but limited to a fewer candidates of multi-component alloys (MCAs) due to the lack of accurate potentials. Therefore, machine learning potentials or force fields trained from the relatively small data sets obtained from the highly accurate quantum mechanical calculations can be a suitable alternative, with an accuracy of \emph{ab initio} methods and efficiency over order of magnitudes than the DFT. Besides, the MLIPs, once trained, interpolate over a wide range of material compositions without requiring repeated DFT calculations for all of them,  which enables high-throughput search of novel MCAs with desired properties. 

Because of their computational efficiency and accuracy, MLIPs have recently gained wide attention in various branches of science~\cite{unke2021machine, musil2021physics, deringer2021gaussian,botu2017machine,ramprasad2017machine}. Most machine learning potentials (MLPs) comprise of two parts: 1) the functional form called the "descriptors", which describes the local chemical environment satisfying the rotations, translations, and reflections symmetries, as well as the permutation of chemically equivalent elements. 2) A regressor, which maps the local environment to the potential energy and its derivative. In 2007, Behler and Parrinello introduced an artificial neural network (NN) as a regressor for MLIP development, which has recently been adopted in various studies~\cite{PhysRevLett.98.146401,behler2016perspective,smith2017ani,pun2019physically}. Bartok \emph{et al.} used Gaussian process regression to build their MLIP, which is called Gaussian Approximate Potentials (GAP)~\cite{PhysRevLett.104.136403}. Alternatively, spectral neighbor analysis potential (SNAP) and moment tensor potential (MTP) are based on linear regression with a set of basis functions~\cite{thompson2015spectral, shapeev2016moment, novikov2020mlip}. 

In this work,  we have used the MTP-based approach, as implemented in the MLIP package~\cite{novikov2020mlip}. The MTPs models are successfully trained for quaternary MoNbTaW (R4) and quinary MoNbTaTiW (R5) refractory high-entropy alloys (HEAs). Unlike conventional alloys, refractory HEAs have high melting point and high strength at elevated temperatures, which make them a suitable candidates for the structural applications~\cite{miracle2017critical, han2017effect}. However, refractory HEAs are brittle at room temperature, which limits their broader applicability~\cite{senkov2010intermetallics}. On the other hand, HEAs can have a non-equimolar combination of principle elements and do not restrict the addition of minor elements~\cite{miracle2017critical,gao2016high}, allowing for a myriad of unexplored alloys with prominent properties. There have been some studies that show better properties of non-equimolar compositions compared to their equimolar counterparts~\cite{li2017trip,li2017strong}. 

The use of MLIP in optimizing the HEA properties is nominal. Most of the machine learning based predictive models for alloys are based on data available in literature or DFT calculated properties   ~\cite{zhou2019machine,chen2021developing, bhandari2021deep}. Such data sets are very small and are unlikely to accurately predict a wide range of properties for the vast configuration space of these alloys. Only recently, MLIPs have been used for efficient optimization of HEA properties ~\cite{hart2021machine,grabowski2019ab,meshkov2019sublattice}. In this work, we demonstrate the application of MLIPs for high-throughput screening of novel HEAs that exhibit promising hardness-ductility combinations, which addresses the hardness-ductility trade-off in HEAs~\cite{li2017strong, lei2018enhanced}. This screening is carried out in for cases: (1) the non-equimolar compositions of R4 and (2) the Ti alloying in R4 (or R5). The MLPs are used for statistical averaging over numerous equimolar and non-equimolar configurations to adequately describe the systems for which DFT calculations alone would be extremely time-consuming and computationally intractable.

The training data sets are generated from the DFT calculations, which are described in the method section below.  The trained models or MLIPs are incorporated in the well-known MD package LAMMPS for high-throughput elastic and mechanical properties calculations.

\begin{figure}
	\centering
		\includegraphics[scale=0.15]{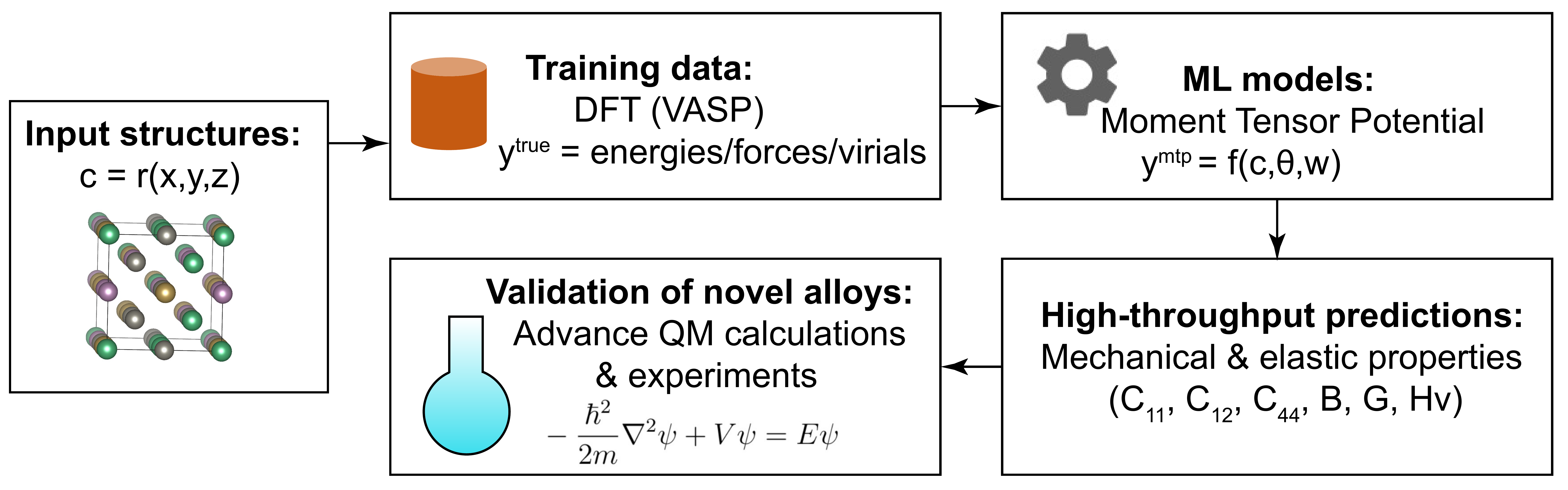}
	\caption{Schematic representation of the workflow for training the MTP interatomic potential for performing high-throughput screening of high-entropy alloys for superior mechanical and elastic properties. }
	\label{work_flow}
\end{figure}
\section{Methods}\label{sec2}
\subsection{Computational methods}
Moment tensor potential (MTP) describes the atomic contribution of an atom \emph{i} to the total energy as:
\begin{equation}
    V_{atom}^{MTP} (\mathbf{r_{i}}) = \sum_{j=1}^{m} \theta_{j} B_{j} (\mathbf{r_{i}})
\end{equation}
where B$_{j}$ are the pre-defined basis functions, \emph{m} is the number of functions in the basis, and $\theta$ are the fitting parameters. The total energy of a configuration is given by:
\begin{equation}
    E_{Total}^{MTP} (c)=\sum_{i=1}^{N_{atm}} V_{atom}^{MTP} (\mathbf{r_{i}})
\end{equation}
where N$_{atm}$ is the total number of atoms in a configuration \emph{c}. The functional form of the basis function B$_{j}$ is defined by the moment tensor descriptors:

\begin{equation}
    M_{\alpha,\beta} (\mathbf{r})= \sum_{j}f_{\alpha}( \lvert \mathbf{r_{ij}}\rvert,z_{i},z_{j})\mathbf{r_{ij}^{1}}\otimes\cdots \otimes \mathbf{r_{ij}^{\beta}}
\end{equation}

where f$_{\alpha}$ is the radial part and \textbf{r$_{ij}^{1}$}$\otimes \cdots \otimes$\textbf{r$_{ij}^{\beta}$} is the angular part of the moments. The functional form and other details of the moments are discussed in \cite{shapeev2016moment,novikov2020mlip}. Then, the force acting on the j$^{th}$ atom is given by:
\begin{equation}
    f_{j}^{MTP}(c)=-\nabla_{x_{j}}E_{Total}^{MTP} (c)
\end{equation}
where x$_{j}$ is the atomic position. Similarly, the virial, the volume weighted stress components are given by,
\begin{equation}
    \sigma^{MTP} (c)= \frac{1}{|det(a)|}(\nabla_{a}E_{Total}^{MTP}(c))a^{T}
\end{equation}
where \emph{a} is the lattice vector. Then, the value of $\theta_{j}$ are determined by minimization of the cost functional (L) given by:
\begin{equation}
   L= \sum_{c\in C_{all}}\lbrack w_{E}^{2}\Delta E(c)^{2}+w_{f}^{2}\sum_{j=1}^{N_{atm}}\Delta f_{j}(c)^{2}+w_{\sigma}^{2}\Delta \sigma (c)^{2}\rbrack
\end{equation}
where, C$_{all}$ is total configurations, $\Delta E$= $E^{MTP} -E^{DFT}$ is the error in total energy. Similarly, $\Delta f$ and and $\Delta \sigma$ are the errors in forces and virials, respectively. $w_{E}$, $w_{f}$, and $w_{\sigma}$ are the weight factors for total energy, force, and stress, respectively.
%
We have implemented the workflow illustrated in Fig.\ref{work_flow} for training the MPT interatomic potential.

The training and validation data sets to fit the MTP potentials are obtained from the quantum mechanical calculations within the density functional theory (DFT)framework. DFT calculations were performed using the VASP package~\cite{kresse1996efficient}. The electron-ion interactions are described by a projector augmented wave (PAW)~\cite{blochl1994projector}, and the generalized gradient approximation of Perdew-Burke-Ernzerhof is used as an exchange-correlation function~\cite{perdew1996generalized}. The DFT calculations for quaternary MoNbTaW (R4) are carried out using a 32-atom 2$\times$2$\times$4 bcc supercell with randomized atoms. For quinary MoNbTaTiW (R5), a 40-atom 2$\times$2$\times$5 bcc supercell with a randomized atom is used as a starting structure. The data are generated in two ways: 1) from \emph{ab initio} molecular dynamics (AIMD) sampling for 20000 steps (20 ps) at a time step of 1fs,  2) from the single point (SP) energy, forces, and virial calculations for random alloys. For AIMD, 400 eV cut-off energy and the k-mesh of 2$\times$2$\times$1 are taken for both R4 and R5. The configurations after every 50 steps are subjected to SP calculations with stringent cut-off energy of 500 eV and k-mesh of 6$\times$6$\times$3 for both R4 and R5. For the SP calculation of random alloys, the cut-off energy of 500 eV and k-mesh of 6X6X3 is taken for all the cases. The AIMD simulations are carried out at temperatures 500K, 1000K, and 1500K with a 2\% increase and decrease in the lattice parameters for each temperature. The random alloys for SP calculations comprise alloys sampled with random compositions, volume (2\% increase and decrease in volume), and shape (2\% increase and decrease in lattice angle). While training the MTP model, 800 configurations were randomly chosen from AIMD and SP generated configurations.  The static and elastic properties are calculated in LAMMPS~\cite{plimpton1995fast} with the trained interatomic potentials. 

For bcc cubic phase, there are three independent single-crystal elastic constants: C$_{11}$, C$_{12}$, and C$_{44}$. For the supercell model we adopted, there are nine independent elastic constants due to low symmetry and large cell size~\cite{wen2019computation}. Therefore, we took the average of the calculated elastic constant as follows:
\begin{equation}
\begin{split}
    C_{11}=\frac{C_{11}^{'}+C_{22}^{'}+C_{33}^{'}}{3}\\
    C_{12}=\frac{C_{12}^{'}+C_{23}^{'}+C_{13}^{'}}{3}\\
    C_{44}=\frac{C_{44}^{'}+C_{55}^{'}+C_{66}^{'}}{3}
\end{split}
\end{equation}
where the elastic constants on the right-hand side are the ones obtained from calculations. The elastic constants are calculated from the optimized structures from the stress-strain relationship method as implemented in the VASP and LAMMPS. The Voigt-Reuss-Hill (VRH) approximation~\cite{Andrews_1978} is used to calculate the bulk modulus (B) and shear modulus (G) of the polycrystalline materials from the elastic constants C$_{ij}$ of a single crystal. B and G in VRH approximation is given by~\cite{wu2007crystal,haines2001annu}:
\begin{equation}
    \begin{split}
        B_{VRH}=\frac{B_{V}+B_{R}}{2}, \\
        G_{VRH}=\frac{G_{V}+G_{R}}{2}
    \end{split}
\end{equation}
where B$_{V}$ and G$_{V}$ is Voigt’s bulk modulus and shear modulus, and B$_{R}$ and G$_{R}$ are Reuss’s bulk modulus and shear modulus, respectively. For cubic phase, they are given by:
\begin{equation}
\begin{split}
    B_{V}=B_{R}=\frac{C_{11}+2C_{12}}{3},\\
    G_{V}=\frac{C_{11}-C_{12}+3C_{44}}{5},\\
    G_{R}=\frac{5(C_{11}-C_{12})C44}{4C_{44}+3(C_{11}-C_{12})}.
\end{split}
\end{equation}
The mechanical stability criteria as suggested by Born and Huang are given by~\cite{waller1956dynamical}:

\begin{equation}
    C_{11}>0, C_{44}>0, C_{11}>|C_{12}|, (C_{11}+2C_{12})>0
\end{equation}

The ductility of materials is estimated through Pugh’s criterion: the condition for the ductility of materials is that the Pugh’s ratio (B/G) should be greater than 1.75~\cite{pugh1954xcii}. Also, the Cauchy pressure (C$_{11}$-C$_{44})$ is positive in a ductile material and negative in the brittle material~\cite{nguyen2008dislocation}. The microhardness is calculated using the Cheng \emph{et al.} model for Vicker's hardness which is based on Pugh modulus ratio k=B/G and is given by~\cite{chen2011hardness}:
\begin{equation}
    H_{V}=2(k^{2}G)^{0.585}-3
\end{equation}
\begin{table}[h]
\begin{center}
\begin{minipage}{170pt}
\caption{ Parameters for MTP model.}\label{tab1}%
\begin{tabular}{@{}lll@{}}
\toprule
Parameter& Value  & Value  \\ 
              & (MoNbTaW) & (MoNbTaTiW)  \\
\midrule
     Radial functions & 4  &5 \\ 
     Radial basis size & 8 & 8   \\ 
     alpha moments count & 350 & 1352 \\ 
     alpha index basic count & 130 & 295 \\
     alpha index times count & 969 & 6349 \\
     alpha scalar moments & 92 & 288 \\ 
     Cutoff & 5.0 \AA  & 5.0 \AA  \\ 
     Stress weight & 1$\times$ 10$^{-3}$  & 1$\times$ 10$^{-3}$  \\ 
     Force weight & 1$\times$ 10$^{-2}$ & 1$\times$ 10$^{-2}$ \\ 
     Energy weight & 1 & 1\\
     BFGS iterations & 1000 & 1000  \\ 
\botrule
\end{tabular}
\end{minipage}
\end{center}
\end{table}
%
%
\begin{table}[h]
\begin{center}
\begin{minipage}{\textwidth}
\caption{Bulk properties of 2X2X4 supercell (32-atom) of equimolar MoNbTaW and 2X2X5 supercell (40-atom) supercell of MoNbTaTiW obtained from the SQS. The properties are calculated using DFT and MPT  at 0K.}\label{tab2}
\begin{tabular*}{\textwidth}{@{\extracolsep{\fill}}lcccc@{\extracolsep{\fill}}}
\toprule%
& \multicolumn{2}{@{}c@{}}{MoNbTaW (SQS)} & \multicolumn{2}{@{}c@{}}{MoNbTaTiW (SQS)} \\\cmidrule{2-3}\cmidrule{4-5}%
      & DFT  & MTP & DFT  & MTP \\
\midrule
 C11 (GPa) & 371.29   & 356.70 (3.9\%) &297.27 &  288.17 (3.06\%) \\ 
 C12 (GPa) &  159.93  & 165.72 (-3.6\%)  &149.49 & 158.15 (-5.8\%)  \\ 
 C44 (GPa) &  75.36  & 68.00 (9.8\%) & 54.99 & 42.43 (22.8\%) \\ 
 VRH Bulk Modulus (B) (GPa) &  230.38  & 229.38 (0.43\%) &198.75 &  201.49 (-1.4\%) \\ 
 VRH Shear Modulus (G) (GPa) & 86.31   & 77.92 (9.72\%) & 61.91 & 50.37 (18.63\%)  \\ 
 Cauchy Pressure (C11-C44) (GPa) &  84.57  & 97.72 & 94.49 & 115.72   \\ 
 Vicker's Hardness (GPa)&  5.60  & 4.23 & 2.71 & 0.91  \\ 
 Pugh's Criterion (B/G) & 2.67   & 2.94  & 3.21&  4.00 \\ 
\botrule
\end{tabular*}
\end{minipage}
\end{center}
\end{table}

\subsection{Experimental methods}
The R4 and R5 alloys were purchased from Sophisticated Alloys. 
All specimens were ground to a 4000 grit SiC paper and further polished to a final solution of 0.04 $\mu$m silica.  A Thermo-Fisher G4 plasma focused ion beam/scanning electron microscope was used to produce elemental maps of the HEAs.  Nanoindentation tests, used to measure the hardness and modulus of the HEAs, were performed on a Keysight G200 Nanoindenter equipped with a diamond Vicker's tip.  All indents were made to a final displacement of 2,000 nm with a constant strain rate (loading rate divided by the load) of 0.05 s$^{-1}$. Continuous stiffness measurements (CSM) were performed at a frequency of 45 Hz and 2 nm displacement amplitude. Hardness and modulus measurements were determined using the Oliver-Pharr method~\cite{oliver2004measurement}. The tip area function was calibrated by indenting fused silica and using tip properties with a Young’s modulus and Poisson’s ratio of 1130 GPa and 0.07 (diamond).

\begin{figure}
	\centering
		\includegraphics[scale=0.3]{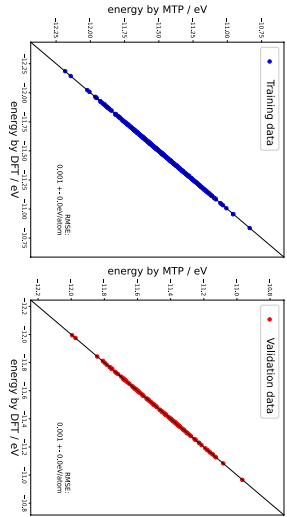}
	\caption{Comparison of total energy of quaternary MoNbTaW alloys with different compositions from DFT and MTP predictions, shown for training (left) and validation (right) data sets. }
	\label{pred_egy_r4}
\end{figure}
\begin{figure}
	\centering
		\includegraphics[scale=0.35]{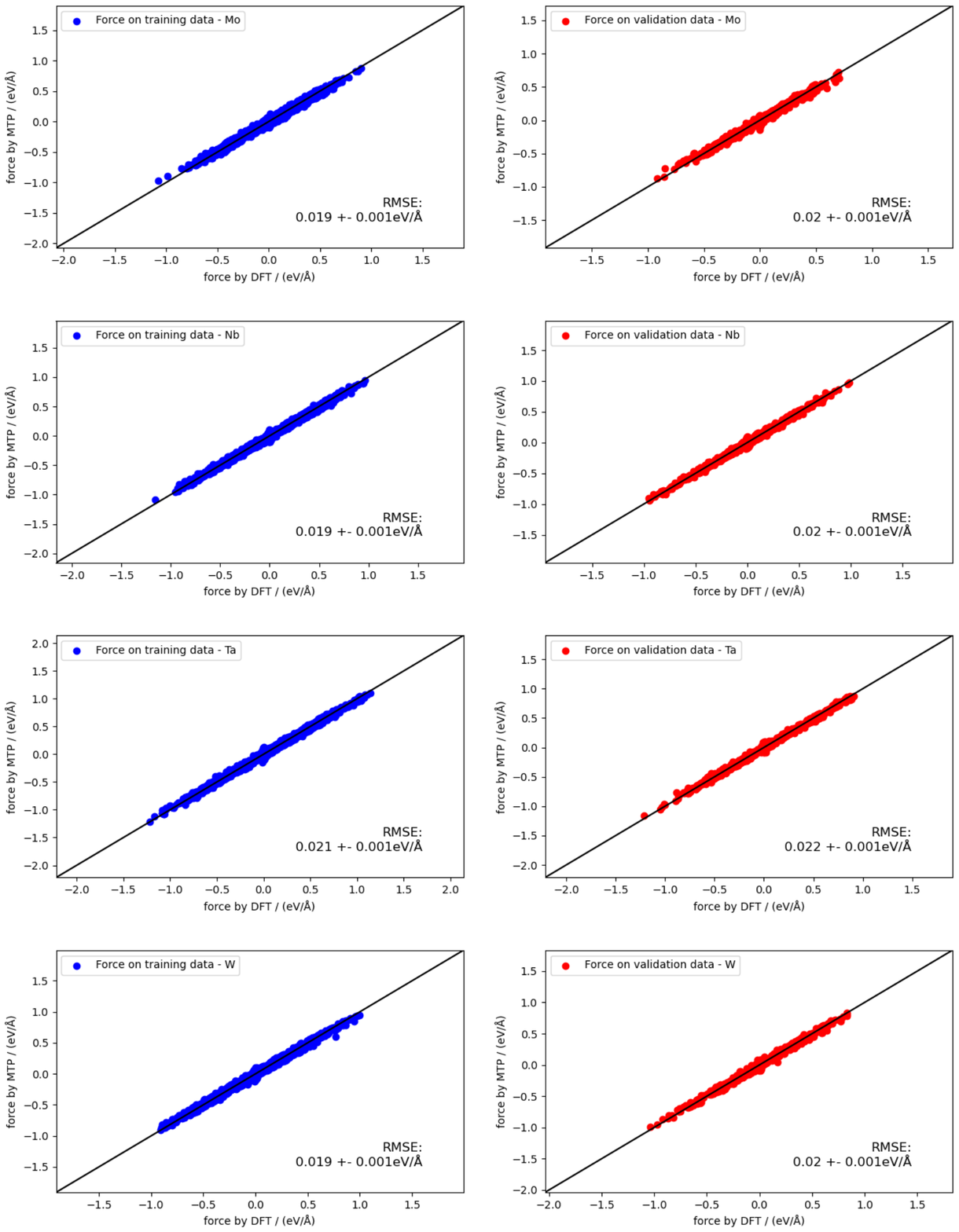}
	\caption{Comparison of forces on quaternary MoNbTaW (both equimolar and non-equimolar compositions) from DFT calculations (ground truth) and MTP predictions  on each element for the training (left) and validation (right) data sets. }
	\label{pred_force_r4}
\end{figure}
\begin{figure}
	\centering
		\includegraphics[scale=0.38]{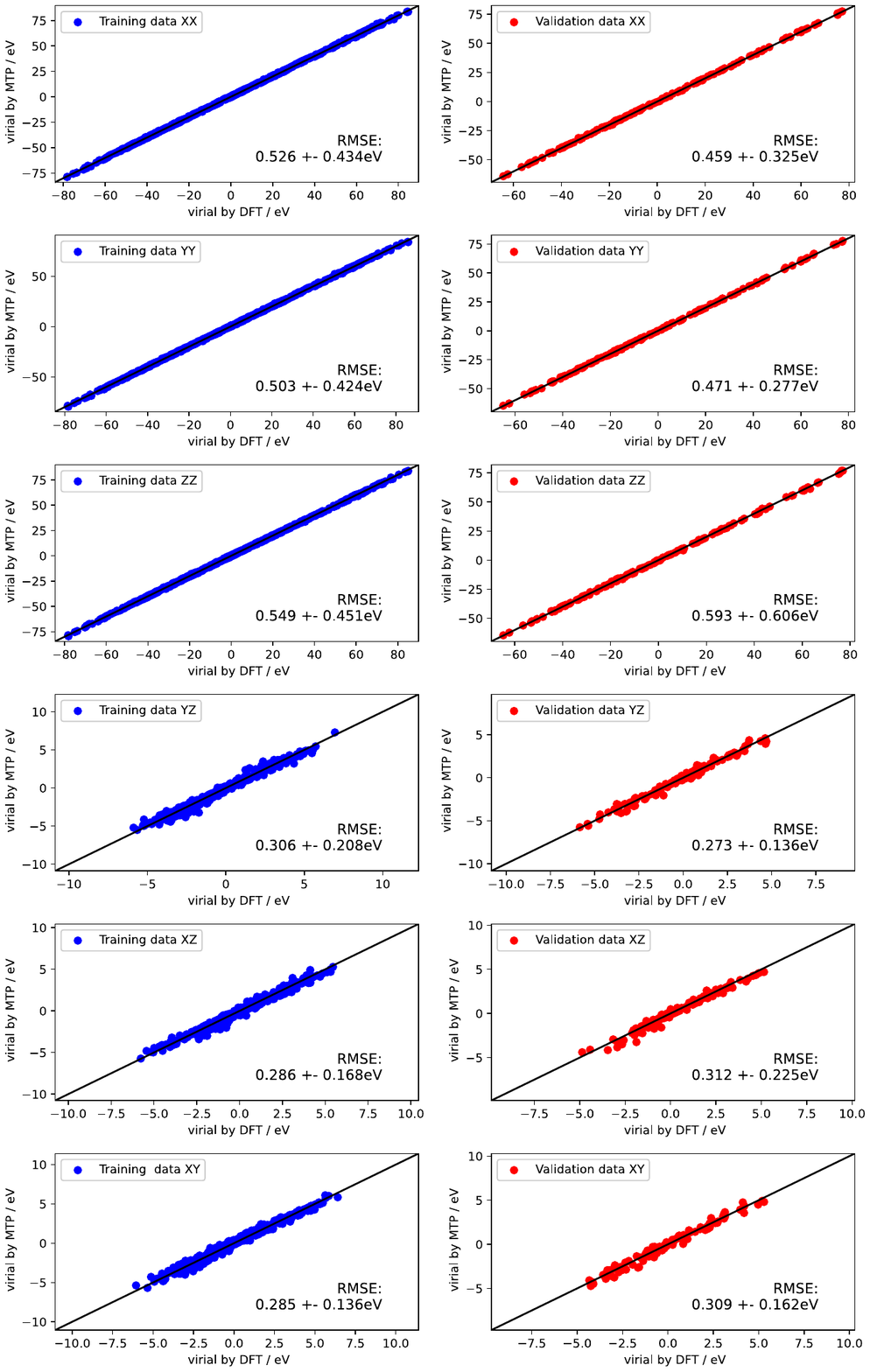}
	\caption{Comparison of six virial components on quaternary MoNbTaW (both equimolar and non-equimolar compositions) from DFT calculations (ground truth) and MTP predictions for training (left) and validation (right) data sets. }
	\label{pred_vir_r4}
\end{figure}

\begin{table}[h]
\begin{minipage}{\textwidth}
\caption{Bulk properties of equimolar MoNbTaW at 0K obtained from MTP by averaging over 50 randomly ordered 432-atom system alloys. The results are compared with the DFT results averaged over 5 random alloys.}\label{tab3}%
\begin{tabular}{@{}llllll@{}}
\toprule
& a (\AA) & B (GPa) & C11 (GPa)& C12 (GPa) & C44 (GPa)  \\
\midrule
MTP &  3.241$\pm$0.0010 & 231.12$\pm$0.51 & 355.39 $\pm$1.01 & 168.97$\pm$0.42 & 69.60$\pm$0.37  \\ 
DFT&  3.241$\pm$0.0002  & 231.44$\pm$0.3  & 375.15 $\pm$0.85&159.58$\pm$0.76 & 74.98$\pm$3.23\\

\botrule
\end{tabular}
\end{minipage}
\end{table}
\section{Results and Discussion}\label{sec3}
The optimized parameters for the trained MTP models for the two different alloy systems are shown in Table \ref{tab1}.

\subsection{MTP for quaternary MoNbTaW}
\begin{figure}
	\centering
		\includegraphics[scale=0.4]{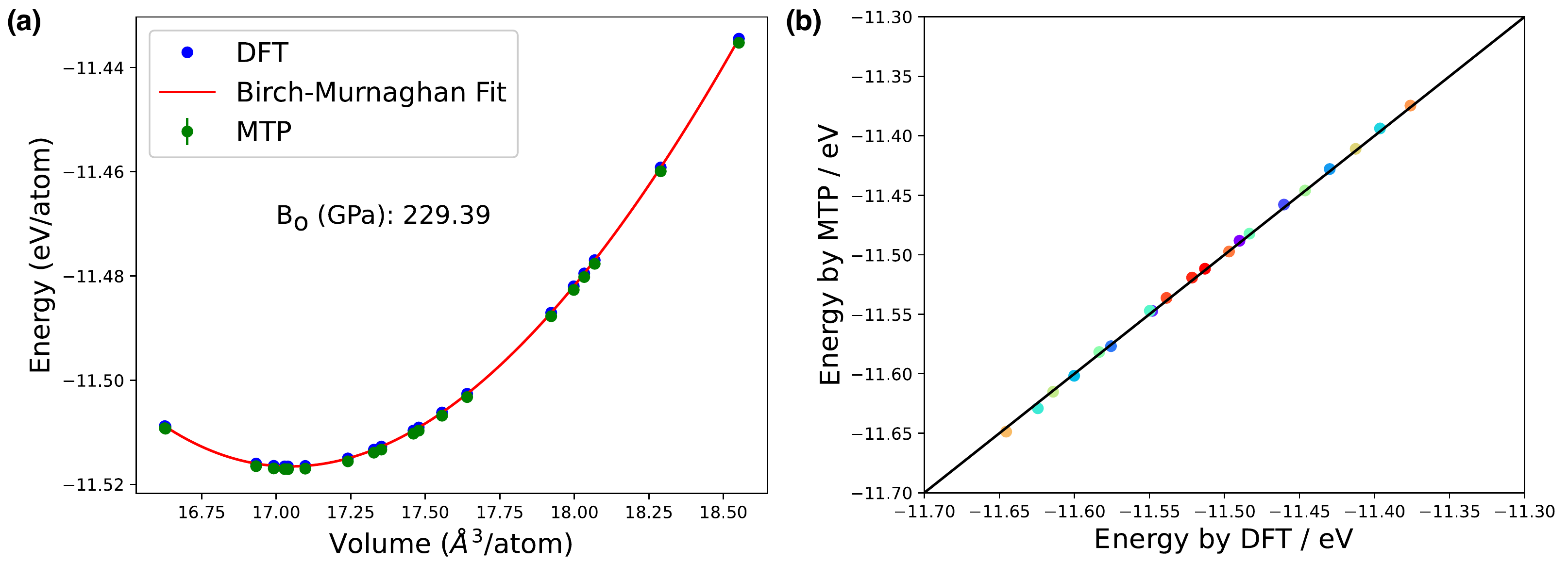}
	\caption{(a)Comparison of energy per atom as a function of volume for an equimolar MoNbTaW random alloy obtained from MTP and DFT. Solid red line is the Birch-Murnaghan fit to the DFT data. (b) Comparison of energy per atom for 20 random 64-atom non-equimolar MoNbTaTiW obtained from MTP and DFT. Colors correspond to different structures.}
	\label{ev_egy_comp}
\end{figure}

The training history for quaternary MoNbTaW (R4) is shown in Fig.\ref{pred_egy_r4}, Fig.\ref{pred_force_r4}, and Fig.\ref{pred_vir_r4}. 
It can be seen that the MTP predicted energy, force, and virial compare well with the DFT calculated values (ground truth). The accuracy for MTP is  within 0.041 eV for energies, 0.021 eV/{\AA} for forces, and 0.383 eV for virials for the validation data. This verifies that the potential is accurate for any given composition of MoNbTaW. The equation of state at 0K for an equimolar R4 obtained from MTP is compared to the DFTs in Fig.\ref{ev_egy_comp}(a). The energy-volume (EV) curve compares well with the DFT, suggesting that the ML-based potential can describe the elastic response accurately. The bulk modulus obtained by fitting the EV data from DFT to the Birch-Murnaghan equation of state~\cite{birch1947finite} is 229.39 GPa. We have adopted a special quasi-random supercell (SQS)~\cite{zunger1990ferreira} (implemented in the icet package $https://icet.materialsmodeling.org/advanced\textunderscore topics/sqs\textunderscore generation.html$) method to generate a 2$\times$2$\times$4 supercell (32-atom) of equimolar R4. The MTP MLIP is further validated by calculating the elastic constants and related properties for the SQS model by incorporating the MTP in the LAMMPS package  and comparing  with the corresponding results from the DFT as shown in Table \ref{tab2}. The elastic moduli predicted by MTP are in excellent agreement with the DFT with errors $\textless$5\% except for a slight overestimation of C$_{44}$ (\% error 9.8\%) and corresponding shear modulus (\% error 9.72\%). The  predictions are better compared to the one from the SNAP machine learning model, where the percentage errors on C$_{11}$, C$_{12}$, C$_{44}$, B, and G are 5.8\%, 3.8\%, 15.9\%, 4.3\% and 13.3\%,  respectively~\cite{li2020complex}.  The Vicker's hardness for the SQS model from the DFT and MTP model are 5.60 GPa and 4.23 GPa, respectively. These values are in a good agreement with the experimental value 4.455 GPa~\cite{senkov2010intermetallics}. The strained SQS models are not included in the training data, yet the MTP model is robust enough to perform well for the R4 SQS model, which is the strong validation of generalization of the MTP model.

We have experimentally measured nonoindentation Vicker's hardness and modulus for MoNbTaW, which are 8.06$\pm$0.31 GPa and 214.4$\pm$2.6 GPa, respectively. Usually, the nanoindentation measurements are overestimated compared to other indentation techniques~\cite{zong2006indentation}. The nanoindentaion bulk modulus compares really well with the DFT calculated value of 230.38 GPa (6.9 \% error). 

In order to ensure the statistical reproduciblity, we have computed average bulk properties of equimolar MoNbTaW from 50 random 432-atom systems using the MTP, which is shown in Table \ref{tab3}. Such type of statistical sampling is not easily achievable by the DFT. The results from MTP are compared with the DFT results averaged over 5 random 32-atom systems. The lattice constant calculated from both the MTP and DFT are 3.241{\AA}, which is close to the experimental value 3.213{\AA} ~\cite{senkov2010intermetallics} and the theoretical value 3.271{\AA}~\cite{li2018first}. 
\begin{figure}
	\centering
		\includegraphics[scale=0.33]{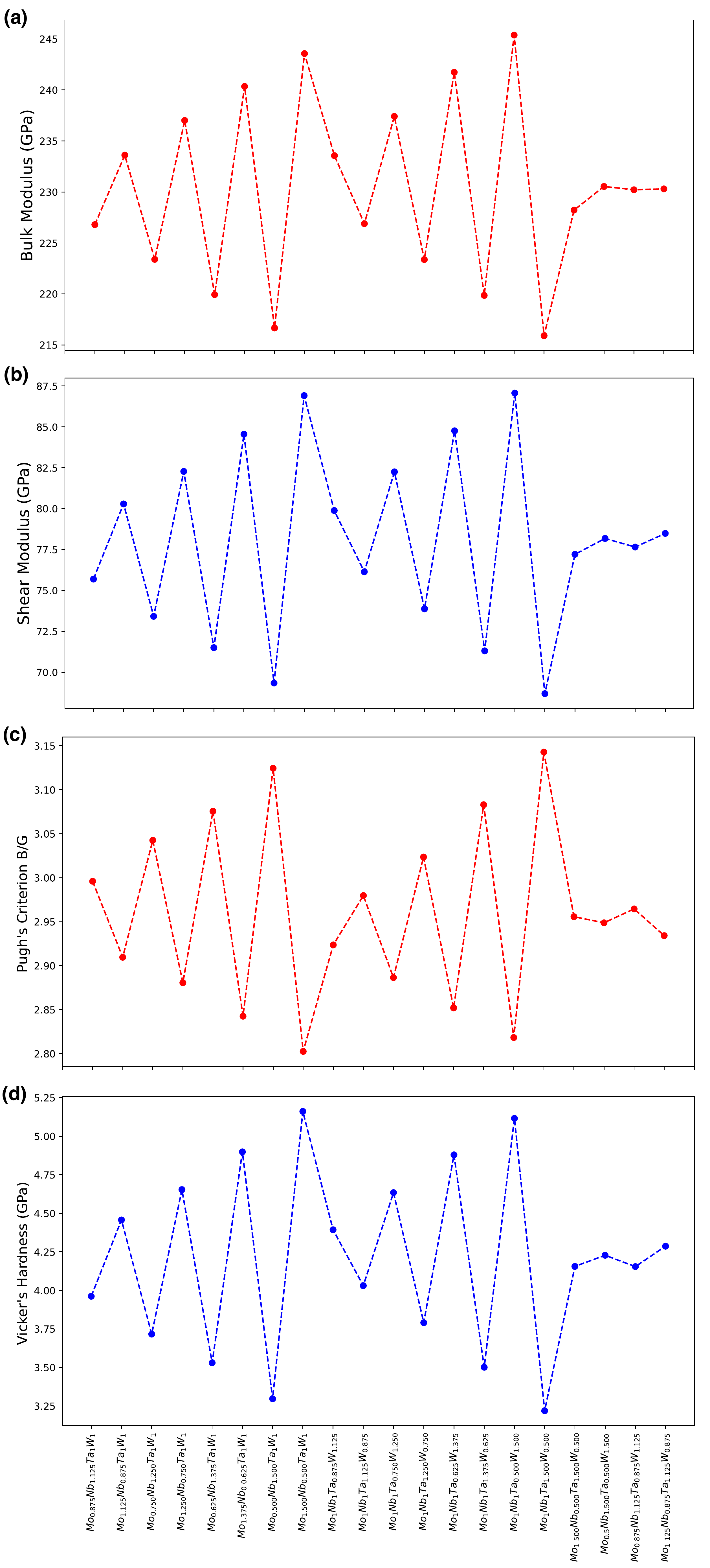}
	\caption{(a)Bulk modulus (b) shear modulus (c) Pugh's Criterion, and (d) Vicker's Hardness for the twenty random 64-atom non-equimolar MoNbTaTiW obtained from MTP. }
	\label{bm_sm_pc_vh_20str_r4}
\end{figure}

\subsection{Rational design of non-equimolar MoNbTaW}
The configurational space for non-equimolar MoNbTaW spans over several orders of magnitudes. It is computationally impossible to screen the components for optimal properties using the highly expensive  DFT calcualtions, before carrying out the expensive experiment. As a test case, we have tweaked the compositions of adjacent atomic number elements pair, Mo-Nb pair and Ta-W pair, one at a time by 12.5\%, 25\%, 37.5\%, and 50\% and screened their elastic response using the MTP. Also, we tested four cases of mixed variations in all four components by 25\% and 50\%. The MTP predicted total energies for twenty non-equimolar structures are in excellent agreement with the DFT, as shown in Fig.\ref{ev_egy_comp} (b) The bulk and shear modulus of the twenty non-equimolar structures are shown in Fig.\ref{bm_sm_pc_vh_20str_r4}(a) and Fig.\ref{bm_sm_pc_vh_20str_r4}(b). The Pugh's criterion and Vicker's hardness are shown in Fig.\ref{bm_sm_pc_vh_20str_r4} (c) and Fig.\ref{bm_sm_pc_vh_20str_r4} (d). 

It can be seen that reducing the Mo composition and increasing the Nb composition lowers the hardness and improves the ductility (Mo$_{0.875}$Nb$_{1.125}$TaW, Mo$_{0.75}$Nb$_{1.25}$TaW, Mo$_{0.625}$Nb$_{1.375}$TaW, and Mo$_{0.5}$Nb$_{1.5}$TaW ). On the other hand, increasing the Mo compositions and decreasing the Nb composition increases the hardness and decreases the ductility (Mo$_{1.125}$Nb$_{0.875}$TaW, Mo$_{1.25}$Nb$_{0.75}$TaW, Mo$_{1.375}$Nb$_{0.625}$TaW, and Mo$_{1.5}$Nb$_{0.5}$TaW ). Similarly, increasing Ta compositions and reducing W decreases the hardness and improves the ductility (MoNbTa$_{1.125}$W$_{0.875}$, MoNbTa$_{1.25}$W$_{0.75}$, MoNbTa$_{1.375}$W$_{0.625}$, and MoNbTa$_{1.5}$W$_{0.5}$). However, reducing the Ta compositions and increasing the W compositions improves the  hardness and reduces the ductility (MoNbTa$_{0.875}$W$_{1.125}$, MoNbTa$_{0.75}$W$_{1.25}$, MoNbTa$_{0.625}$W$_{1.375}$, and MoNbTa$_{0.5}$W$_{1.5}$). There is no significant variation in the hardness and ductility when all four compositions are varied in a mixed way (Mo$_{1.5}$Nb$_{0.5}$Ta$_{1.5}$W$_{0.5}$, Mo$_{0.5}$Nb$_{1.5}$Ta$_{0.5}$W$_{1.5}$, Mo$_{0.875}$Nb$_{1.125}$Ta$_{0.875}$W$_{1.125}$, and Mo$_{1.125}$Nb$_{0.875}$Ta$_{1.125}$W$_{0.875}$). 

Some non-equimolar R4 candidates with promising hardness-ductility combinations, for example Mo$_{0.75}$Nb$_{1.25}$TaW (B/G= 3.04, H$_{V}$=3.72), Mo$_{0.625}$Nb$_{1.375}$TaW (B/G= 3.07, H$_{V}$=3.53), MoNbTa$_{1.25}$W$_{0.75}$ (B/G= 3.02, H$_{V}$=3.79), and MoNbTa$_{1.375}$W$_{0.625}$ (B/G= 3.08, H$_{V}$=3.50), than the equimolar R4 (B/G= 2.94, H$_{V}$=4.23) are predicted by the MTP. The predicted alloy candidates can be experimentally designed and characterized~\cite{lamsal2019transmission,Lamsal:20,lamsal2020maximizing,an2020high} for further validation.   

%

\begin{figure}[h]
	\centering
		\includegraphics[width=0.9\textwidth]{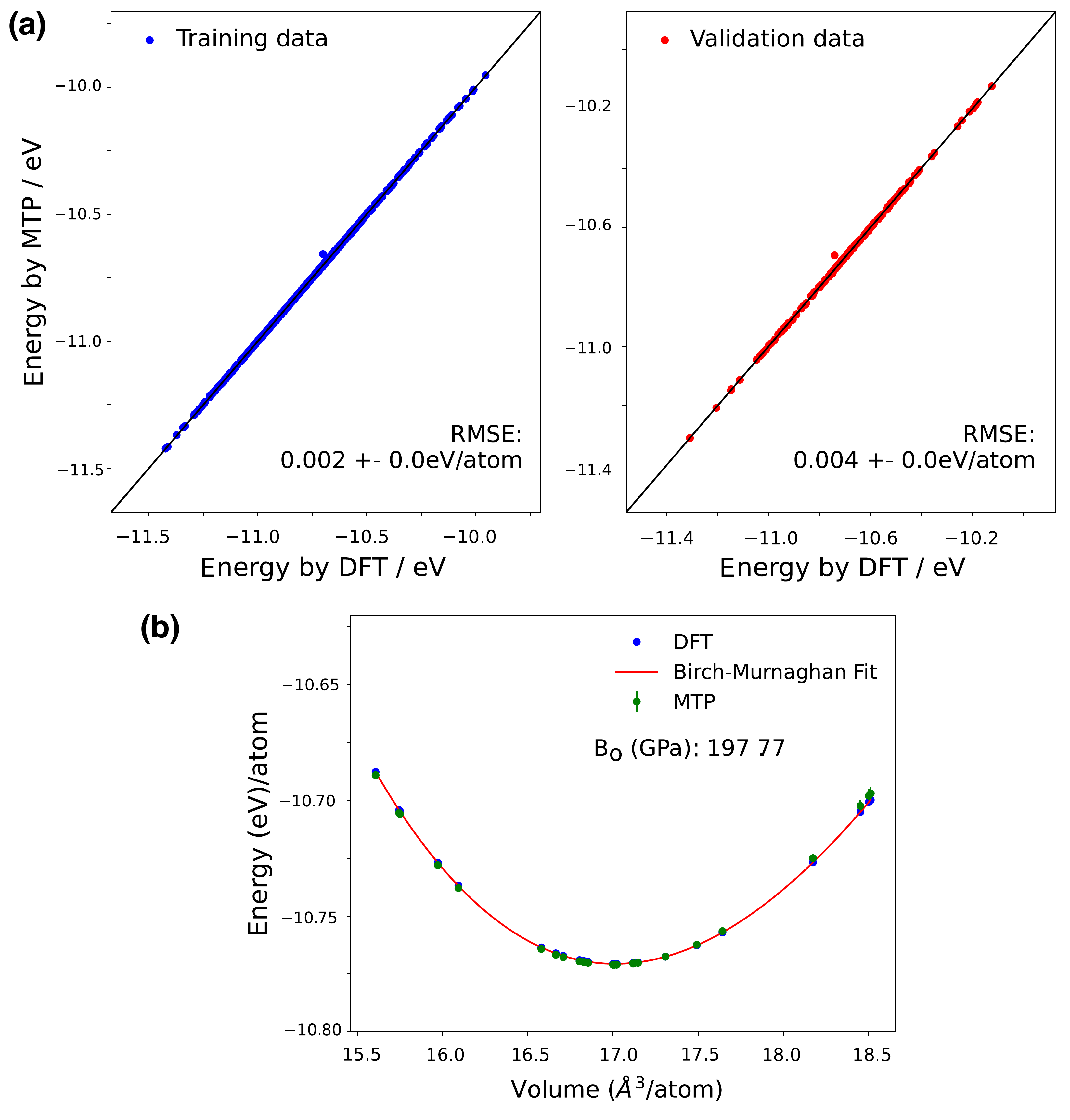}
	\caption{(a) Comparison of total energy of quinary MoNbTaTiW alloys with different compositions from DFT and MTP predictions, shown for training (left) and validation (right) data sets. (b) Comparison of energy per atom as a function of volume for an equimolar MoNbTaTiW random alloy obtained from MTP and DFT. Solid red line is the Birch-Murnaghan fit to the DFT data.}
	\label{pred_egy_ev_r5}
\end{figure}

\begin{figure}
	\centering
		\includegraphics[width=0.9\textwidth]{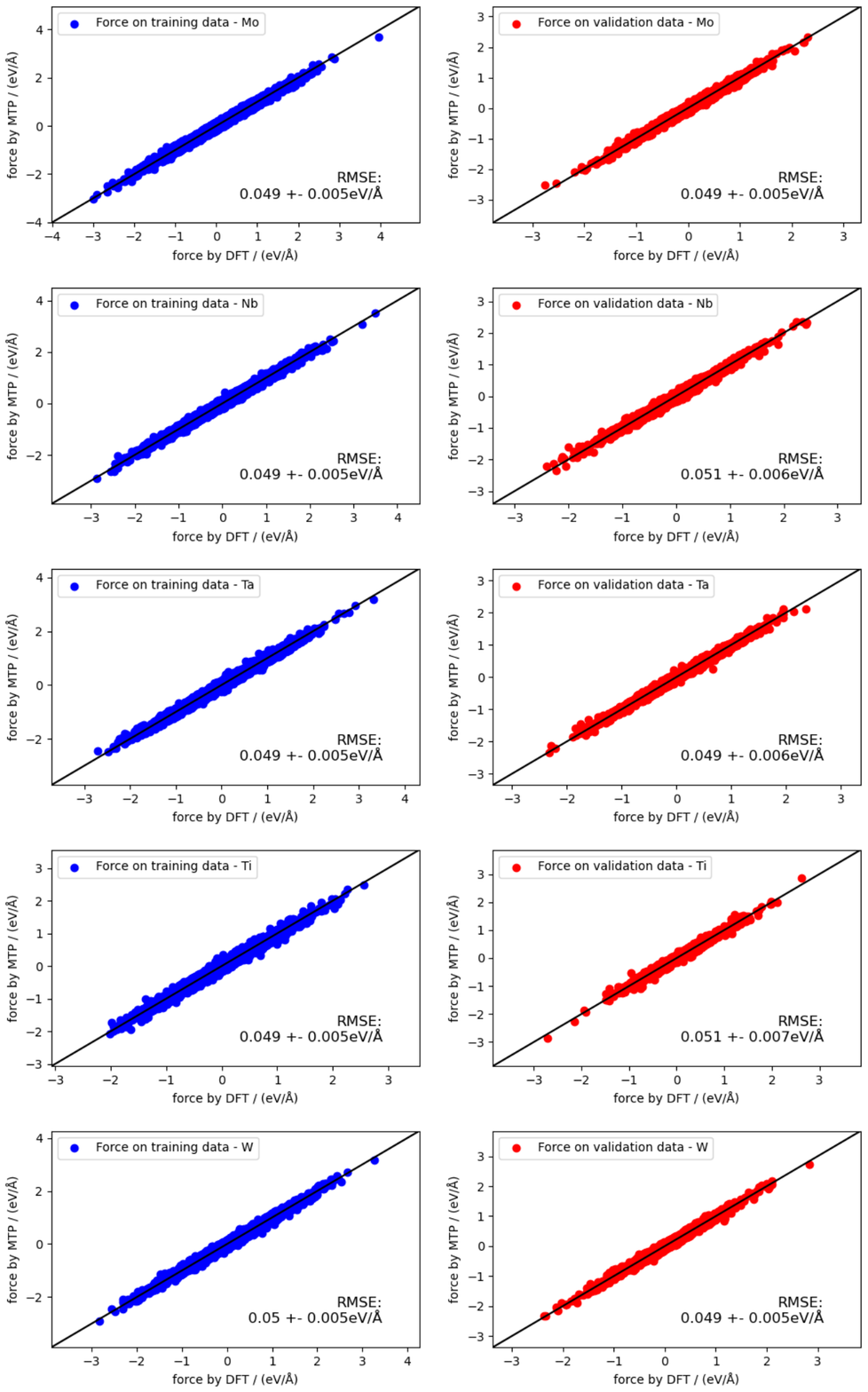}
	\caption{Comparison of forces on quinary MoNbTaTiW (both equimolar and non-equimolar compositions) from DFT calculations (ground truth) and MTP predictions  on each component for training (left) and validation (right) data sets. }
	\label{pred_force_r5}
\end{figure}
\begin{figure}
	\centering
		\includegraphics[scale=0.38]{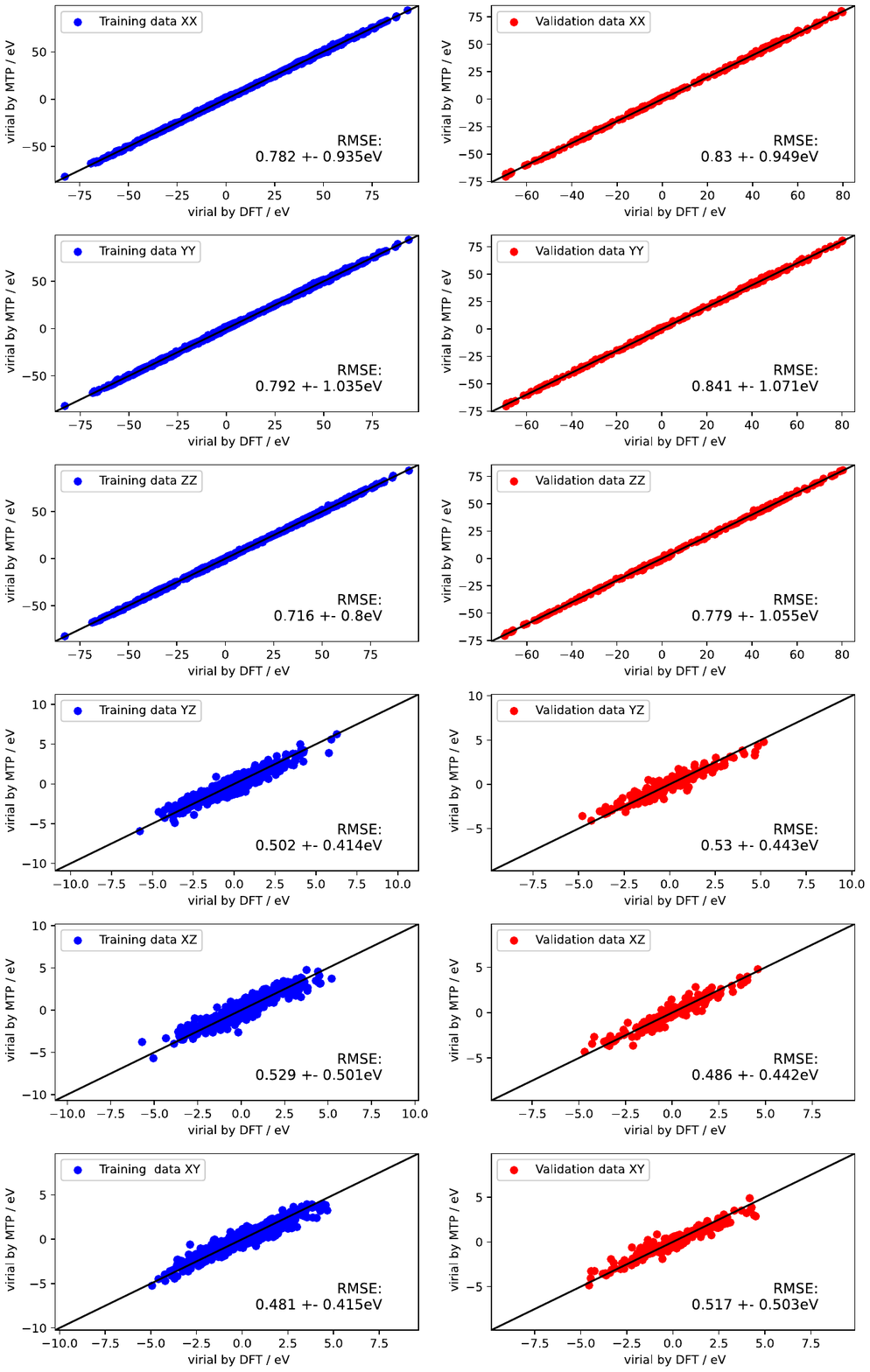}
	\caption{Comparison of six virial components on quinary MoNBTaTiW (both equimolar and non-equimolar compositions) from DFT calculations (ground truth) and MTP predictions for training (left) and validation (right) data sets. }
	\label{pred_vir_r5}
\end{figure}
%
\begin{table}[h]
\begin{center}
\begin{minipage}{\textwidth}
\caption{Bulk properties prediction for random MoNbTaTi$_{0.5}$W. The properties are calculated using DFT and MTP  at 0K. The values from DFT are averages over 5 random structures and that from MTP are averaged over 100 random 180-atom structures }\label{tab4}
\begin{tabular*}{\textwidth}{@{\extracolsep{\fill}}lcc@{\extracolsep{\fill}}}
\toprule%
& \multicolumn{2}{@{}c@{}}{MoNbTaTi$_{0.5}$W}  \\
\midrule
     & DFT & MTP  \\ 
     \midrule
     C11 (GPa)& 319.91$\pm$5.92   & 323.14$\pm$1.61 (-1.01\%)  \\ 
     C12 (GPa)&  158.15$\pm$2.16  & 159.18$\pm$0.43 (-0.65\%)  \\ 
     C44 (GPa)&  55.44$\pm$6.84  & 44.20$\pm$0.49  (20.27\%)   \\ 
     VRH Bulk Modulus (B) (GPa)&  212.07  & 213.83 (-0.83\%) \\ 
     VRH Shear Modulus (G) (GPa) & 64.47   & 56.75 (11.97\%)  \\ 
     C11-C44 (GPa)&  264.47 & 114.99   \\ 
     Vicker's Hardness (Hv) & 2.68   & 1.50 \\ 
     Pugh's Criterion (B/G) & 3.29   & 3.78   \\ 
\botrule
\end{tabular*}
\end{minipage}
\end{center}
\end{table}
\begin{figure}
	\centering
		\includegraphics[scale=0.25]{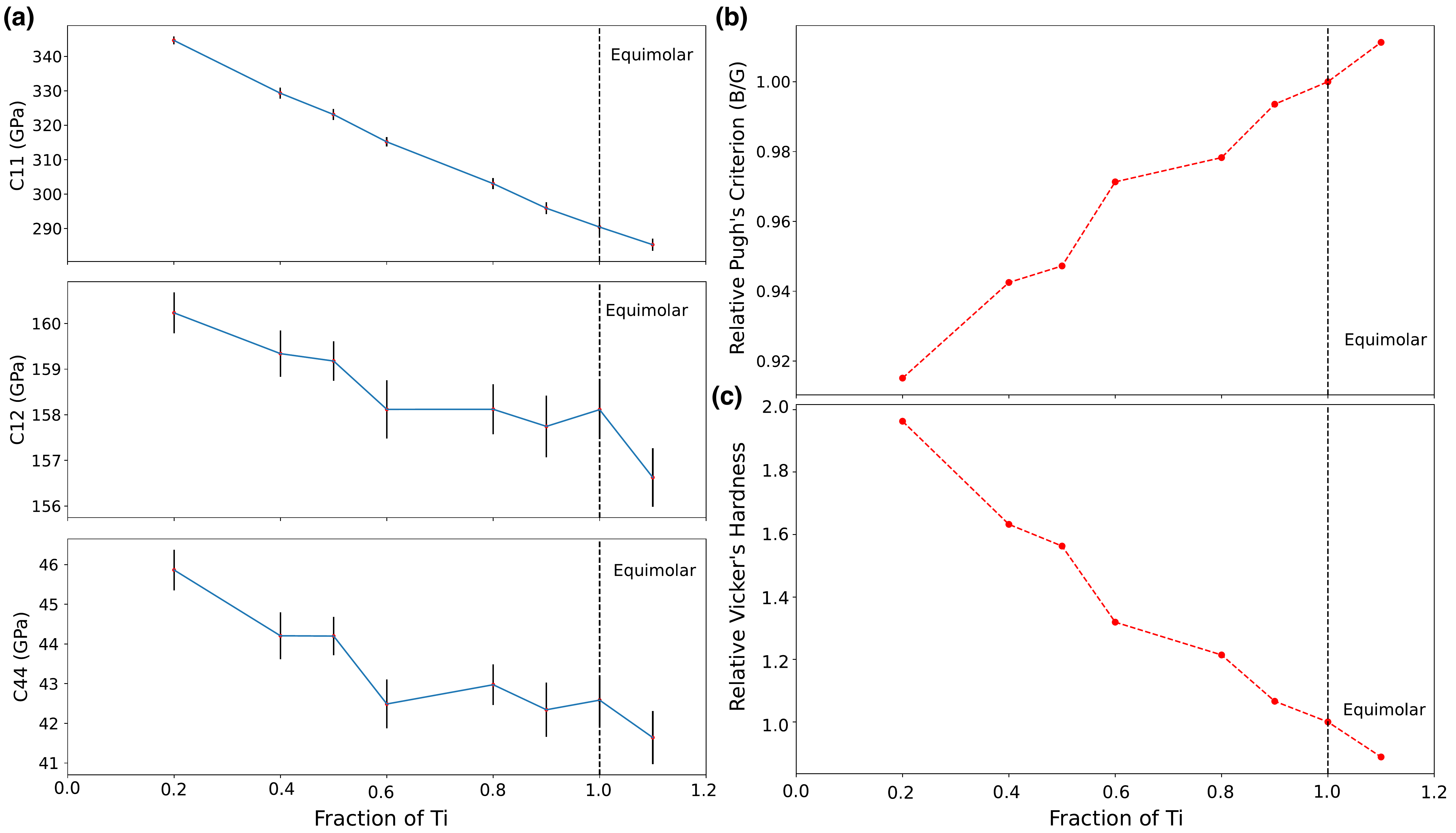}
	\caption{(a) Components of elastic constants for MoNbTaTi$_{x}$W (x=0.2, 0.4, 0.5, 0.6, 0.8, 1.0, and 1.1),  obtained from MTP by averaging over 100 random structures for each composition.  Error bars indicate the the standard deviations for 100 structures at each composition. (b) Pugh's criterion (B/G)) relative to the equimolar MoNbTaTiW, and (c) Vicker's hardness relative to the equimolar MoNbTaTiW obtained from MTP for the nine different Ti concentrations.  }
	\label{ec_pc_vh_ti_sampling_r5}
\end{figure}
\subsection{Alloying effect of Ti on MoNbTaW}

Although it is well established that the mechanical properties can be significantly improved by alloying, experimentally it is not feasible to fabricate and test the alloys with multiple possible compositions in a systematic way. Therefore, it is imperative to have a tool that can provide a high-throughput screening of the various fraction of the alloying element for a desired property and can guide the experiments. This section will demonstrate the applicability of MLIP in the high-throughput investigation of alloying effect of Ti in the elastic and mechanical properties of R4. 
\subsubsection{MTP for quinary MoNbTaTiW}
We have trained a separate MTP with the training data described in the computational method section above. The training parameters are shown in Table \ref{tab1}.
The training history for quinary MoNbTaTiW (R5) is shown in Fig.\ref{pred_egy_ev_r5}(a), Fig.\ref{pred_force_r5}, and Fig.\ref{pred_vir_r5}. The predicted energy, force, and virial compare well with the DFT results (ground truth). The accuracy of MTP is  within 0.151 eV for energies, 0.050 eV/{\AA}  for forces, and 0.630 eV for virials for the validation data, which verifies that the potential is accurate for any given composition of MoNbTaW. As in the case of R4, the equation of state at 0K for an equimolar R5 obtained from MTP compares well with the DFTs as shown in Fig.\ref{pred_egy_ev_r5}(b), suggesting that the ML potential can describe the elastic response accurately. The bulk modulus obtained by fitting the energy-volume data of the DFT to the Birch-Murnaghan equation of state~\cite{birch1947finite} is 197.77 GPa. As in R4, a 2$\times$2$\times$5 supercell (40-atom) SQS model of equimolar R5 is generated. The MTP potential is further validated by calculating the elastic constants and bulk  properties for the SQS model and comparing with the corresponding DFT results as shown in Table \ref{tab2}. The elastic moduli predicted by MTP are in excellent agreement with the DFT with errors $\textless$5\%, except for a slight overestimate for the C$_{44}$ (\% error 22.8\%) and corresponding shear modulus (\% error 18.63\%). Again, the strained SQS models are not included in the training data, yet the MTP model is robust enough to perform well for SQS R5, which is the strong validation of generalization of the MTP model. The Vicker's hardness for the SQS model from the DFT and MTP model are 2.71 GPa and 0.91 GPa, respectively. The difference in the Vikers's hardness is due to the error in the predicted shear modulus. 

 We have experimentally measured nanoindentation Vicker's hardness and bulk modulus for MoNbTaTiW, which are 7.58$\pm$0.30 GPa and 197.3$\pm$7.20 GPa, respectively.The experimentally measured bulk modulus, shear modulus, and the Vicker's hardness  of R5 as reported by Han \emph{et al.} are 139 GPa, 59GPa, and 4.89 GPa, respectively ~\cite{han2017effect}. The nanoindentaion bulk modulus compares really well with the DFT calculated value of 198.75 GPa (0.73 \% error).

In this work, the DFT calculated bulk modulus and shear modulus for the SQS model is 198.75 GPa and 61.91 GPa, respectively. Mishra \emph{et al.} calculated the bulk modulus and shear modulus for the R5 using supercell models and the DFT, which are reported as 193.90 GPa and 70.67 GPa, respectively~\cite{mishra2019}. Similarly, Bhandari \emph{et al.} calculated B and G using the  Knuth-shuffle supercell model and the DFT, which are 199 GPa and 60 GPa, respectively~\cite{bhandari2020first}. This shows that the theoretical results for various models are pretty consistent. There is a considerable discrepancy in the theoretically reported Vicker's hardness and the experimentally measured ones~\cite{han2017effect}, which calls for the further experimental investigations.

\subsubsection{Statistical study of alloying effect }

We have computed the average bulk properties of MoNbTaTi$_{0.5}$W from 100 random 180-atom systems using the MTP, which is shown in Table 4. The results are compared with the DFT calculated values averaged over 5 random 36-atom systems. The elastic moduli predicted by the MTP are in excellent agreement with the DFT with errors around 1\%, except for the C$_{44}$ (\% error 20.27\%) and corresponding shear modulus (\% error 11.97\%). The \% error for C$_{44}$ (and the corresponding shear modulus) for MoNaTaTiW and the MoNbTaTi$_{0.5}$W are consistent, suggesting that the MTP predictions for C$_{44}$ are standardized and can be used in a comparative investigation of other Ti compositions. It should be noted that since the random structures are included in the training sets, the predictions are better than the SQS model. 

The elastic constants predicted for MoNbTaTi$_{x}$W (x=0.2, 0.4, 0.5, 0.6, 0.8, 1.0, and 1.1) are shown in Fig.\ref{ec_pc_vh_ti_sampling_r5} (a), where the error bars are the standard deviations for 100 random structures. The Pugh's ratio and the Vicker's hardness predicted for 9 different Ti concentrations relative to the equimolar R5 are shown in Fig.\ref{ec_pc_vh_ti_sampling_r5}(b) and Fig.\ref{ec_pc_vh_ti_sampling_r5}(c), respectively. The MTP model captures the inverse behavior of ductility (B/G) and the hardness with the addition of the Ti. The addition of Ti reduces the hardness and improves the ductility, which corroborates with the experimental results~\cite{han2017effect}. As shown in Fig.\ref{ec_pc_vh_ti_sampling_r5}, the alloys, for example,  MoNbTaTi$_{0.5}$W  and MoNbTaTi$_{0.8}$W, with Ti composition other than the widely investigated equimolar MoNbTaTiW have promising hardness-ductility combination, which requires further experimental investigations.

We have demonstrated that MTP-MLIP successfully reproduces the \textit{ab initio} potentials of complex high-entropy alloys, at orders of magnitude lower computational cost. The computational efficiency of MTP enabled the screening of vast configurational space occupied by R4 and R5 alloys to optimize their compositions by statistically averaging over hundreds of structures; otherwise not feasible using DFT calculations. Selective tuning of elemental compositions of R4 (Mo, Nb, Ta, W) enabled the discovery of novel refractory HEAs with promising hardness-ductility combinations. This approach was also successfully demonstrated for Ti alloying in R4. Hence, this method can be readily extended for high-throughput screening of other HEA systems and advanced theoretical study of novel alloys for accelerated materials discovery and design. Future work entails quantum mechanical calculations under desired thermo-mechanical conditions and corresponding experimental validation of promising alloy candidates (e.g., Mo$_{0.75}$Nb$_{1.25}$TaW and MoNbTaTi$_{0.5}$W).

MTP-MLIP also exhibited improved accuracy compared to SNAP-MLIP \cite{li2020complex} for the MoNbTaW HEA system studied in this work. Therefore, MTP-MLIP can be used to investigate advanced material phenomena at higher time scales such as the interplay between short-range order, segregation, strengthening, etc., through atomistic simulations,  similar to that reported by Li et al. using the SNAP-MLIP method \cite{li2020complex}. Computational efficiency and the accuracy of DFT also make MTP-MLIP a promising alternative to other ML models trained using material properties data obtained from limited published literature ~\cite{zhou2019machine,chen2021developing}.

There is a considerable discrepancy between the experimentally measured elastic and bulk properties of MoNbTaTiW by Han \emph{et al.} and the DFT calculated ones. Therefore, for further validation, we have measured the Vicker’s hardness and bulk modulus of MoNbTaW and MoNbTaTiW from the nanoindentation technique. The MTP predicted bulk modulus for both the alloys are in excellent agreement with our experimental bulk modulus. However, some inconsistencies exist between the experimentally measured and theoretically calculated Vicker’s hardness~\cite{han2017effect}.

\section{Conclusion}\label{sec4}

We have successfully trained MTP machine learning potentials for quaternary MoNbTaW and quinary MoNbTaTiW HEAs. High-throughput calculations of elastic and mechanical properties of hundreds of novel structures are feasible with MTP-MLIP, which are highly time-consuming using DFT calculations. MTP-MLIPs enabled the rational design of non-equimolar alloys leading to the discovery of novel HEAs with enhanced hardness-ductility combinations (e.g., Mo$_{0.75}$Nb$_{1.25}$TaW). Computationally efficient and accurate MLIPs are reliable predictive tools for high-throughput optimization of alloying elements and screening of other high-entropy alloys. The approach presented here will significantly reduce the exploratory cost when designing a new alloy or optimizing existing ones.

\section{Acknowledgements}\label{sec5}
This work was supported by the US Department of Energy through Los Alamos
National Laboratory. Los Alamos National Laboratory (LANL) is operated by Triad National Security, LLC, for the National Nuclear Security Administration of the US Department of Energy (Contract No. 89233218CNA000001). This work was funded by Institute for Materials Science, Rapid Response Award and LANL’s Laboratory Directed Research and Development Project \#20190571ECR. The experimental work was performed, in part, at the Center for Integrated Nanotechnologies, an Office of Science User Facility operated by LANL. The use of LANL’s HPC institutional computing for the computational time is also acknowledged. We would also like to thank Dr. Ghanshyam Pilania and Dr. Danny Perez for fruitful discussions.

\section{Conflict of Interest}\label{sec6}
The authors declare no conflict of interest.

\bibliography{mybibfile}

\end{document}